\documentclass[a4paper,twocolumn,prl]{revtex4-2}%

\usepackage[latin9]{inputenc}
\usepackage{amsmath}
\usepackage{amssymb}
\usepackage{graphicx}
\usepackage{comment}
\usepackage{mathrsfs}
\usepackage{amsfonts}
\usepackage{float}%
\usepackage{multirow}
\providecommand{\U}[1]{\protect\rule{.1in}{.1in}}
\makeatletter

\DeclareFontEncoding{LGR}{}{}
\DeclareTextSymbol{\~}{LGR}{126}
\@ifundefined{textcolor}{}
{
\definecolor{BLACK}{gray}{0}
\definecolor{WHITE}{gray}{1}
\definecolor{RED}{rgb}{1,0,0}
\definecolor{GREEN}{rgb}{0,1,0}
\definecolor{BLUE}{rgb}{0,0,1}
\definecolor{CYAN}{cmyk}{1,0,0,0}
\definecolor{MAGENTA}{cmyk}{0,1,0,0}
\definecolor{YELLOW}{cmyk}{0,0,1,0}
}

\makeatother
\begin{document}
%\title{Exceptional Flexoelectric Effect in Silicon Thin Film under Bending}
%\title{A New Mechanism of Flexoelectricity Beyond the Linear Response Theory}
%\title{Exceptional Flexoelectric Effect of Silicon: A New Mechanism of Flexoelectricity}
\title{ Transversal Flexoelectricity of  Semiconductor Thinfilm under High Strain Gradient}
%\title{Mechanism of Giant Flexoelectricity in Silicon Thinfilms Beyond the Linear Response Theory}
\author{Chao He$^{1}$}
\author{Jin-Kun Tang$^{1}$}
\author{Yang Yang$^{1}$}
\author{Kai Chang$^{2,3,5}$}
\thanks{Corresponding author}
\email[]{kchang@semi.ac.cn}
\author{Dong-Bo Zhang$^{1,4}$}
\thanks{Corresponding author}
\email[]{dbzhang@bnu.edu.cn}
\affiliation{$^{1}$College of Nuclear Science and Technology, Beijing Normal University, Beijing 100875, P.R. China}
\affiliation{$^{2}$SKLSM, Institute of Semiconductors, Chinese Academy of Sciences, P.O. Box 912, Beijing 100083, China}
\affiliation{$^{3}$CAS Center for Excellence in Topological Quantum Computation, University of Chinese Academy of Sciences, Beijing 100190, China}
\affiliation{$^{4}$Beijing Computational Science Research Center, Beijing 100193, P.R. China}
\affiliation{$^{5}$Beijing Academy of Quantum Information Sciences, Beijing 100193, China}
\begin{abstract}
The flexoelectric behaviors of solids under high strain gradient can be distinct from that under low strain gradient. Using the generalized Bloch theorem, we investigate theoretically the transversal flexoelectric effects in bent MgO thinfilms. As a comparison, a centrosymmetric (100) film and a non-centrosymmetric (111) film are considered. Under bending, the mechanical responses of both films are linear elastic under low strain gradient but nonlinear elastic under high strain gradient. In the linear elastic regime, no internal displacements and thus no polarization contributed from ions are induced. Only in the nonlinear elastic regime, atoms adopt discernibly large  internal displacements, leading to strong polarization from ions. Because the internal displacements of atoms of the (111) film are much larger than those of the (100) film, the obtained flexoelectric coefficient of the (111) film is also greater than that of the (100) film, revealing strong anisotropy of flexoelectricity of MgO film. Our results and the employed approach have important implications for the study of flexoelectric properteis of ionic solids.
\end{abstract}

%\pacs{75.50.Dd, 71.70.Fk, 73.20.Pr, 72.25.Dc}
\maketitle
Recent experimental and theoretical investigations have substantially advanced our knowledge of flexoelectricity~\cite{review1,review2,review3}, a phenomenon that depicts the coupling between the electric polarization and the strain gradient of materials~\cite{Kogan}. Since the crystal symmetries are broken under an inhomogeneous strain, the flexoelectricity is believed to exist in materials with arbitrary symmetry. This is distinct from another phenomenon, piezoelectricity~\cite{martin,Tagantsev}. The latter exists only in non-centrosymmetric crystals. The investigation of flexoelectric effects  has been attracting extensive attention recently. In various materials such as ferroelectric ceramics~\cite{Cross3,Cross4,Catalan1,prl38,prl44} and also in hybrid semiconductors~\cite{Catalan3,Catalan5,wang}, experimentation has revealed strong flexoelectric effects. These advances make it promising for the utility of flexoelectricity in actual applications.

In bulk solids, the studies of flexoelectricity usually involve relatively low level of strain and also low level of strain gradient. For example, in the cantilever bending experiments, the exerted strain gradient is usually at the level of $\sim 0.1$~m $^{-1}$ and the maximal strain in the deformed sample is less than $1\%$~\cite{Cross1,Cross2,surface,prl38}. From the perspective of mechanics, such low strain gradient and strain ensure that the structural deformation of materials are within the linear elastic regime. It is interesting to point out that the low strain gradient is also required by the linear response theory that delineates  the variations of electronic charge density and of the ion displacement due to strain gradient~\cite{Tagantsev5,Resta1,Sharma1,Hong1,Hong2}. The linear response theory not only serves as the microscopic mechanism to delineate the formation of flexoelectricity, but also provides a route to evaluate quantitatively the flexoelectric coefficients. In this aspect, several first-principles approaches  have been developed~\cite{Hong3,prl13,Stengel3}.

However, for materials under high strain gradient,  their flexoelectric behaviors may deviate significantly from the prescribed linear dependence on strain gradient~\cite{wang,pv,prl35,prl20,high1,high2,mgo}. For example, Wang $et$ $al.$~\cite{wang} showed  that a tip indentation can induce an inhomogeneous strain field within a small area of semiconductors where the maximal strain is measured as $\sim5\%$. Consequently, the resulting strain gradient is as high as $10^6$~m$^{-1}$. Due to this high strain gradient, the  flexoelectric response displays a nonlinear behavior that enhances the measured flexoelectric coefficients to a great extent. Mechanically, this odd behavior of flexoelectricity can be attributed to the complex  structural distortion due to high strain gradient, which cannot be delineated directly with the linear response treatment. Theoretically, atomistic simulations are the method of choice to address the issue.

%Except the local structural deformations, high strain gradients can be also achieved in nanoscale systems due to the small length scale, giving rise to strong flexoelectric effects~\cite{Sharma2,prl16,prl22,prl26}. Indeed, it is expected that the role of flexoelectricity may be competitive with that of piezoelectricity at the nanoscale~\cite{review1}.

In this work we investigate theoretically the  flexoelectric response of bent MgO (100) and (111) thinfilms with atomistic quantum mechanical (QM) calculations. For both films, we identify two mechanically elastic regimes of the bending deformations: linear and nonlinear. In the linear elastic regime, we find no internal displacements~\cite{review1,Tagantsev,review2} of atoms, and thus no polarization from ions formed. Only in the nonlinear elastic regime, atoms adopt the internal displacements, leading to non-vanishing polarization. We also reveal that the the internal displacements in the (111) film are much larger than those in the (100) film due to the anisotropy of mechanical rigidity. Consequently, the (111) film has a stronger  flexoelectric effect than the (100) film. This difference illustrates the anisotropic behavior of MgO thinfilm. Our calculations of the bent motifs are enabled by using the generalized Bloch theorem~\cite{gbt1,gbt2,gbt3} coupled with the density-functional tight-binding (DFTB)~\cite{dftb1,dftb2}.

\begin{figure}[tb]
\includegraphics[width=1\columnwidth]{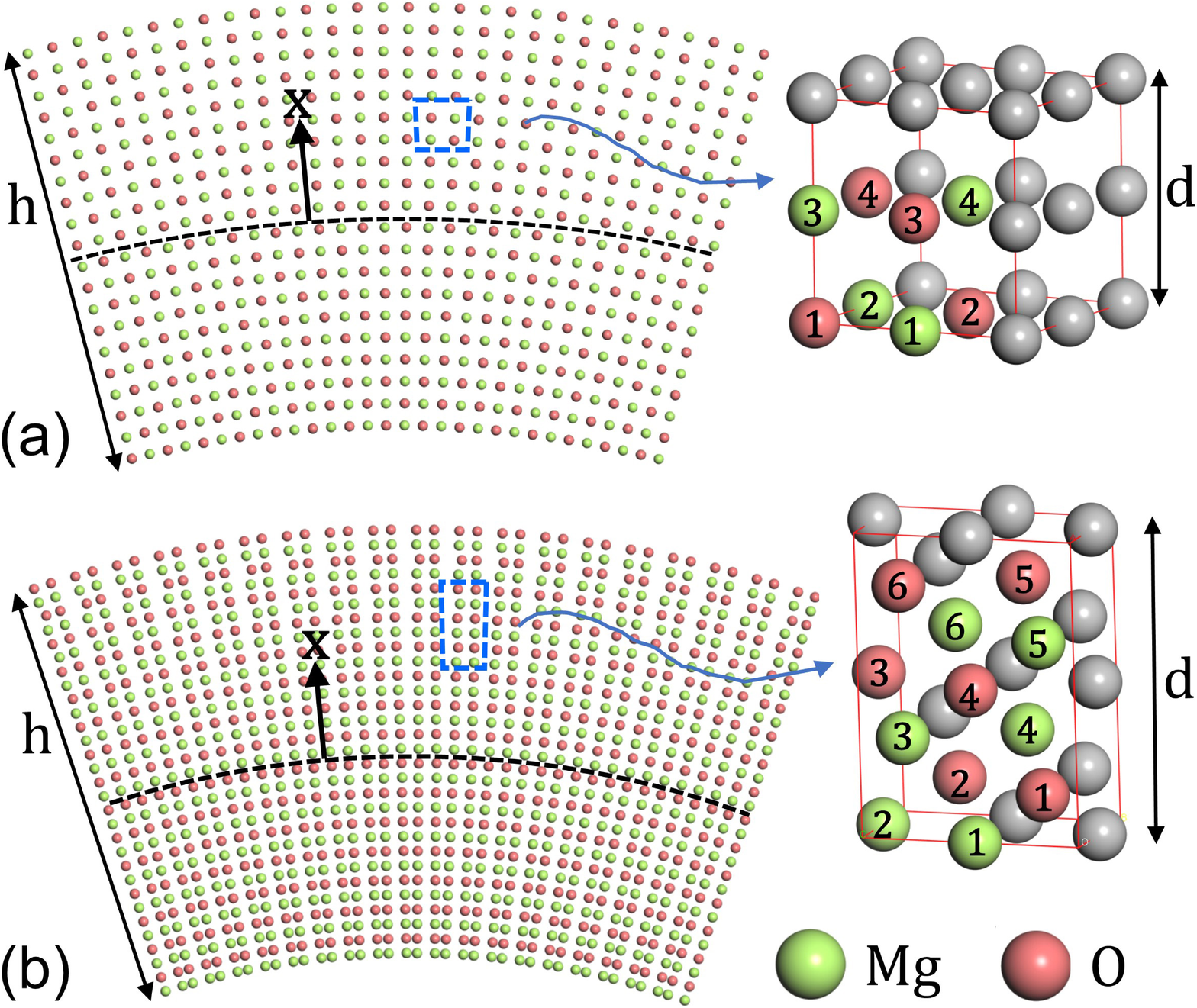}
\caption{ Schematic illustration of  bent MgO thinfilms (left) and structural units (right) for (a) (100) film  and (b) (111) film. $h$ denotes the film thickness and $-h/2<x<h/2$ along the thickness dimension defines the distance from the neutral surface (dashed curve) of the bent film. $d$ measures the unit size along the film thickness dimension. }%
\label{film}%
\end{figure}

Fig.~\ref{film}(a) shows schematically the side view of a bent MgO (100) thinfilm. In order to depict the structural distortion due to bending, we follow the guidance of Cauchy-Born rule~\cite{cauchy} by choosing a structural unit as showcased in Fig.~\ref{film}(a)[right]. In this way, the structural deformation of the bent film can be characterized in terms of the variation of geometrical parameters of this unit and the atomic displacements inside the unit. Note that this unit satisfies the translation symmetry in the [100] crystallographic direction that corresponds to the thickness dimension of the film.  With this structural unit, it is straightforward to realize that the stress-free (100) thinfilm  has a centrosymmetry in the film thickness dimension. This centrosymmetry breaks in the bent film. Inside the unit, atoms are indexed with integer numbers for both Mg and O atoms. For the (111) film, a structural unit can be similarly identified in Fig.~\ref{film}(b). Apparently,  this unit indicates that the stress-free (111) film does not own a centrosymmetry. It is important to further note that the internal displacements of atoms~\cite{review1,Tagantsev,review2} of the bent films in the film thickness direction consist of contributions from the displacements of atoms inside the structural unit and from the variation of the unit length $d$.  By definition, the internal displacements of atoms are the key quantities that determine the induced polarization from ions.

The relaxed configurations of  the bent thinfilms are obtained by carrying out atomistic QM simulations. Because the bending deformation breaks the translational symmetry along the principal curvature. This makes the first-principles and other QM calculations formulated with periodic boundary conditions intractable. Here, the obstacle is overcome  by employing the generalized Bloch scheme. In this scheme, the bent film is described with basic repetition rules of translation ${\bf T}$ and rotation of angle $\theta$ performed in the curvilinear coordinate,
\begin{equation}
\label{bending}
{\bf X}_{\xi,\lambda,n}=\xi{\bf T}+{\bf R}^{\lambda}(\theta){\bf X}_{n},
\end{equation}
where, ${\bf X}_{n}$ represents atoms inside the repeating cell and ${\bf X}_{\xi,\lambda,n}$ represents the atoms inside the replica of the repeating cell indexed by ($\xi,\lambda$).  The repeating cell contains the same $N$ atoms in the translational cell of the undistorted film. For example, for the (100) film with 10~nm thickness, $N=192$. Index $n$ runs over the $N$ atoms inside the cell. The relatively small $N$ allows for systematic QM simulations of the bent structure. The structural relaxation is conducted through a conjugate gradient energy minimization to the repeating cell.

\begin{figure}[tb]
\includegraphics[width=1\columnwidth]{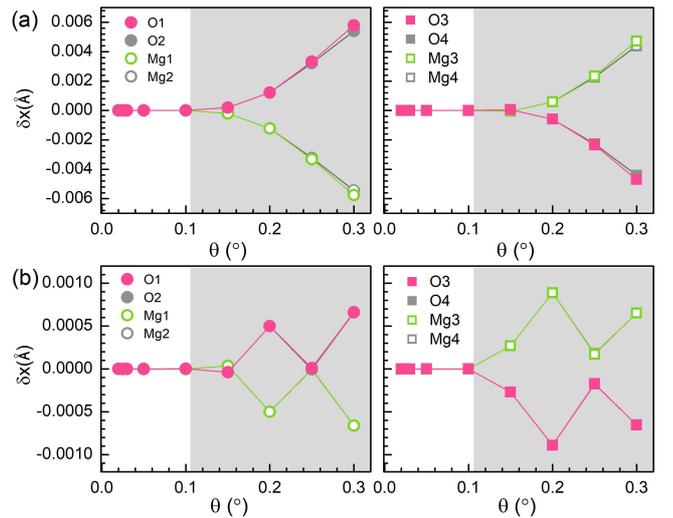}
\caption{ Displacements of atoms inside the structural unit of the bent MgO (100) thinfilm with a 10~nm thickness under different bending angles for the structural unit  centered at (a) $x=-4$~$nm$ on the compressive side and for the structural unit centered at (b) $x=4$~$nm$ on the elongated side.}%
\label{001}%
\end{figure}

Due to the breaking of the centrosymmetry, it is reasonable to expect that atoms of the bent film adopt internal displacements, giving rise to polarization in the strain gradient direction~\cite{review1,review2}. However, we reveal that the bent thinfilm does not necessarily develop internal displacements merely due to symmetry breaking. We focus on the atom displacements in the thickness dimension. For the MgO (100) thinfilm with thickness $h=10$~nm, Fig.~\ref{001} shows that at relatively small bending angle $0<\theta<0.1^{\circ}$ corresponding to strain gradient $0<g<4.13\times10^6$m$^{-1}$, the   displacements of atoms inside the considered structural unit remain vanishingly small, i.e., $\delta x\simeq0$, whether the structural unit is on the compressive side or on the elongated side. From the viewpoint of mechanics, under small $\theta$, the structural distortion is within the linear elastic regime, thus the planer arrangement of Mg and O atoms along the principal curvature is well maintained, Fig.~\ref{film}(a). Indeed, at $0<\theta<0.1^{\circ}$, the considered structural units feel only low strain $0<|\varepsilon|<1\%$.

At $\theta>0.1^{\circ}$ (i.e., $g>4.13\times10^6$m$^{-1}$), the   displacements of atoms are nonzero and reveal several interesting aspects. (i) Mg and O atoms in one atomic layer adopt opposite displacements. (ii) Although Mg atoms in one atomic layer along the principal curvature [Fig.~\ref{film}(a)] are of nearly equal displacements, Mg atoms in different atomic layers have different displacements.  (iii) The   displacements of atoms inside  the structural unit on the compressive side, Fig.~\ref{001}(a),  are much larger than those of the structural unit on the elongated side, Fig.~\ref{001}(b). Roughly speaking, this is because that the atom layers on the compressive side of the bent film are squeezed severely. Large displacements of atoms in the thickness dimension can effectively release the strain energy. On the contrary, the atom layers on the elongated side of the bent film are stretched. As such, the displacements of atoms are not preferred.  These reflects the mechanical response in the nonlinear elastic regimes of the bent film.

\begin{figure}[tb]
\includegraphics[width=1\columnwidth]{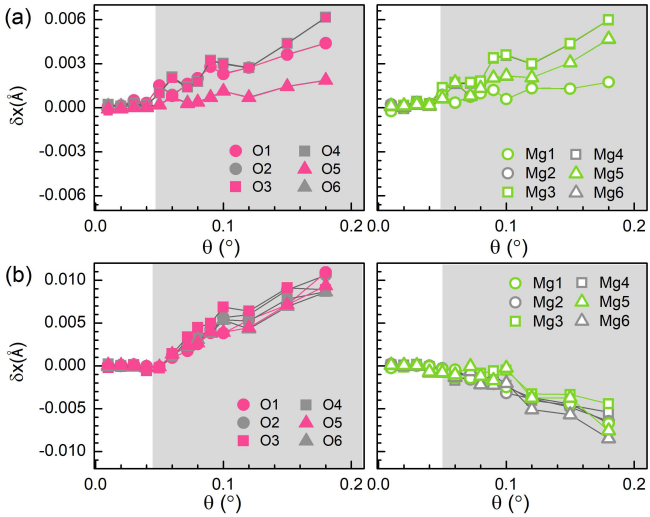}
\caption{ Displacements of atoms inside the structural unit of the bent MgO (111) thinfilm with a 10~nm thickness under different bending angles for the structural unit  centered at (a) $x=-4$~$nm$ on the compressive side and for the structural unit centered at (b) $x=4$~$nm$ on the elongated side. }%
\label{111}%
\end{figure}

Similarly, the mechanical behavior of the MgO (111) thinfilm under bending also undergoes two distinct stages, Fig.~\ref{111}. At small bending angle  $\theta<0.04^{\circ}$ (the corresponding strain gradient $0<g<1.35\times10^6$m$^{-1}$), the structural deformation is within the linear elastic regime, where the displacements along the film thickness dimension $\delta x\simeq 0$ for all the atoms. However, at $\theta>0.04^{\circ}$ ($g>1.35\times10^6$m$^{-1}$), the structural deformation is beyond  the linear elastic regime with nonzero $\delta x$. The atoms inside the considered structural units are of different   displacements unless they are within the same atomic layers along the principal curvature of the bent film, Fig.~\ref{film}(b). What is interesting is that the   displacements of the structural unit on the elongated side [Fig.~\ref{111}(b)] are even greater than those of the structural unit on the compressive side [Fig.~\ref{111}(a)]. This is different from the situation for the (100) film, revealing the mechanical anisotropy of the MgO thinfilm in the nonlinear elastic regime.

\begin{figure}[tb]
\includegraphics[width=1\columnwidth]{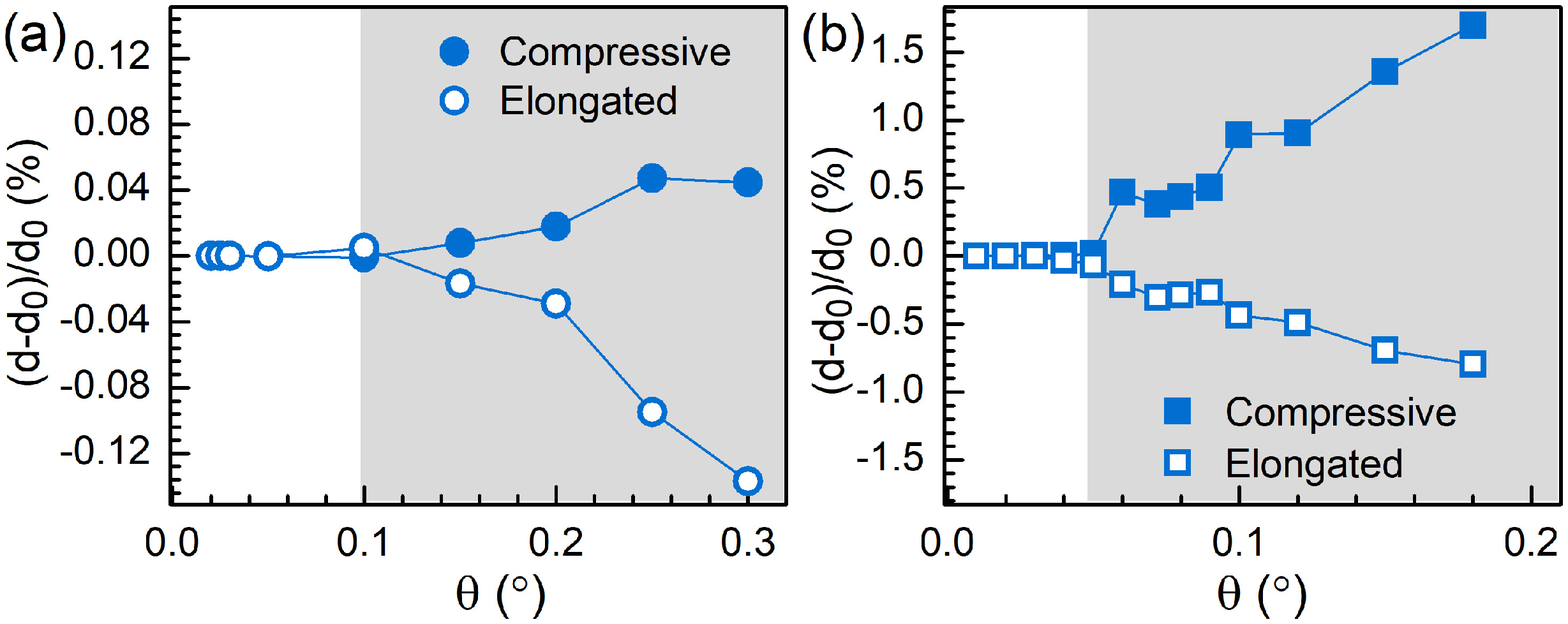}
\caption{Variations of the unit size in the thickness dimension versus bending angle for the structural unit  centered at $x=-4$~$nm$ on the compressive side and for the structural unit centered at $x=4$~$nm$ on the elongated side of the bent films for (a) (100) film and (b) (111) film. Both thinfilms are 10 nm thick.  }%
\label{unit}%
\end{figure}

\begin{figure}[b]
\includegraphics[width=0.6\columnwidth]{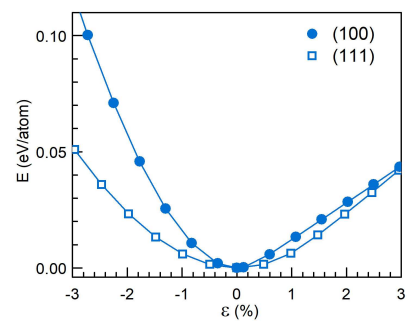}
\caption{Strain energies of MgO versus normal strain for (a) the strain along [100] direction and (b) the strain along [111] direction.  }%
\label{energy}%
\end{figure}

Another dimension to probe the structural distortion due to bending is to observe the shape changes of the structural unit. Here we focus on the unit length ($d$) along the film thickness dimension. Fig.~\ref{unit} displays the percent variations of the unit length for bent (100) and (111) thinfilms, $(d-d_0)/d_0$. Here $d_0$ is the unit length for the stress-free films. For both films, the variations of the unit length with bending angle have two stages. At small bending angle ($0<\theta<0.1^{\circ}$ for the (100) film and $0<\theta<0.04^{\circ}$ for the (111) film), $(d-d_0)/d_0\simeq 0$. This result is consistent with the fact that the bending deformation is within the linear elastic regime for both films. When the bending angle further increases, the situation is different where the variations of the unit length are significant. Specifically, for both bent films,  $(d-d_0)/d_0> 0$ for the unit on the compressive side, while $(d-d_0)/d_0< 0$ for the unit on the elongated side. These asymmetric results reflect the mechanical behavior of solids in the nonlinear regime.

Further, we also notice that the amplitude of $(d-d_0)/d_0$ for the (111) film is much greater than that for the (100) film, which is related with  the anisotropy of mechanical rigidities between different crystallographic directions. Indeed, Fig.~\ref{energy} displaying the strain energy of MgO under normal strains in different directions shows that the strain energy of the strain in [100] direction is higher than that of the strain in [111] direction. This reveals that MgO is more rigid in [100] direction  than in [111] direction. Correspondingly, in the bent MgO thinfilms,  the structural unit in the (111) film is easier to adopt a shape change in terms of $d$  than the structural unit in the (100) film.

%In summary, we have explored the structural distortions of MgO (100) and (111) films due to bending by analyzing the variations of structural units. For both films, their mechanical responses are characterized with a linear elastic behavior at small bending angle and a nonlinear elastic behavior at large bending angle. In the linear elastic regime, the structural units do not adopt any atomic   displacements or variation in unit length. Consequently, it is reasonable to expect that no polarizations are induced. On the contrary, in the nonlinear elastic regime, the structural units develop discernible   displacements and variations in unit length. This result has important implication. For the (100) film, due to the centrosymmetry along the film thickness dimension, the variation of unit length cannot induce polarization. The net polarization thus originates mainly from the   displacements of atoms. For the (111) film, it does not own the centrosymmetry in the thickness dimension. Therefore, both the   displacements of atoms and the variation of unit length have contributions to the polarization.

\begin{figure}[tb]
\includegraphics[width=1\columnwidth]{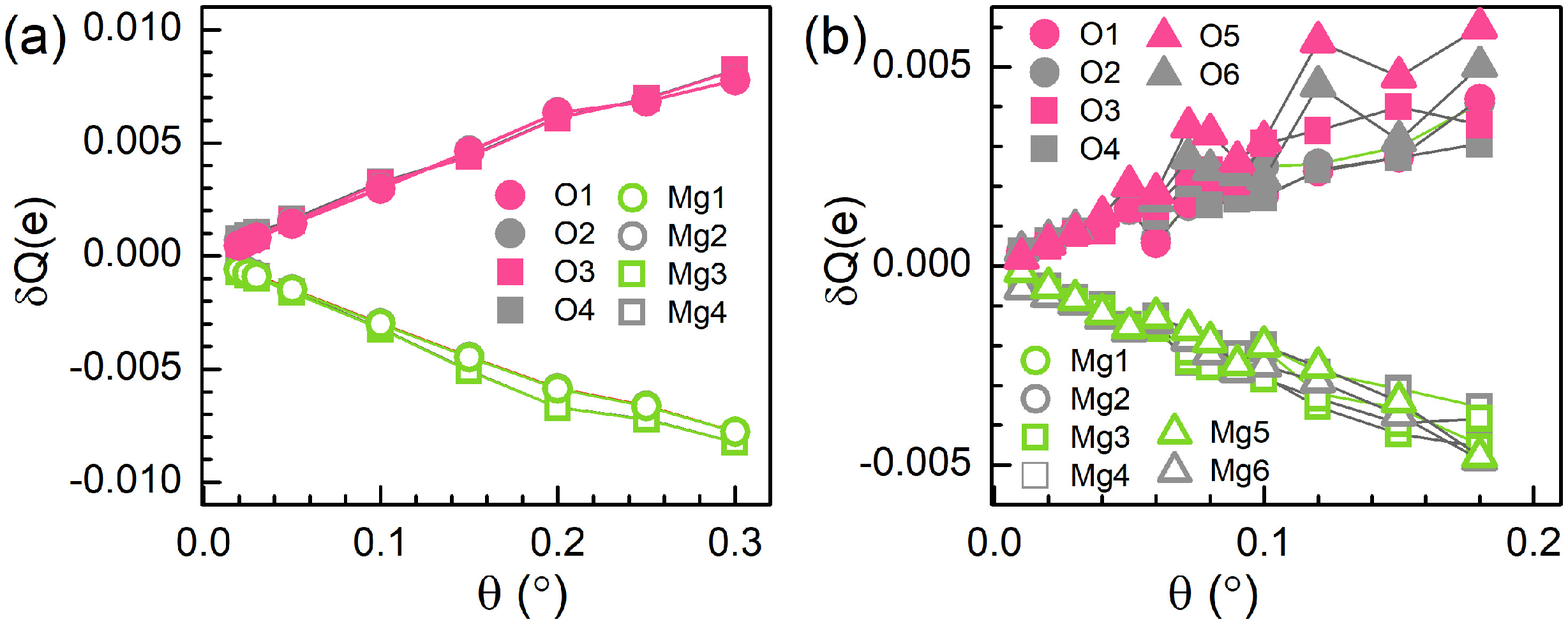}
\caption{Variations of the atomic charge versus bending angle for the structural unit  centered at $x=4$~$nm$ on the elongated side of the bent films for (a) the (100) film and (b) the (111) film. }%
\label{charge}%
\end{figure}

Combining the result of atomic displacements inside the structural unit and the result of the shape variation of the structural unit, we are able to conclude that the internal displacements of atoms in the (111) film are larger than those in the (100) film. Therefore, it is reasonable to expect that stronger polarization can be induced in the bent (111) film than in the (100) film.
In order to gain more insight into the formation of the polarization, we have evaluated the bending induced polarization in the thickness dimension using the structural data as input for both films,
\begin{equation} \label{ppp}
{\bf P}(\theta)=\frac{1}{\Omega}\int_{-h/2}^{h/2} Q(x)xdx,
\end{equation}
where, $\Omega $ denotes the volume of the repeating cell. $Q$ is the atomic charge of the atom at position $x$. The origin of $x$ is at the neutral surface, Fig.~\ref{film}. Notice that the (100) film has the centrosymmetry in the thickness dimension, thus ${\bf P}=0$ for the stress-free case. As such, the bending induced polarization, $\Delta {\bf P}={\bf P}$. On the other hand, the (111) film does not have the centrosymmetry, the stress-free film has a nonzero ${\bf P}$. In this case, the bending induced polarization is calculated as $\Delta {\bf P}={\bf P}({\theta})-{\bf P}(0)$.

A simple way to determine the values of atomic charges is to use the rigid-ion model~\cite{Tagantsev5,review1} where ions are of fixed charges. Here, $Q=2$~($e$) for O and $Q=-2$~($e$) for Mg. However, the atomic charges may vary with the bending deformation due to the charge transfer between Mg and O atoms. For a more realistic treatment, we calculate the atomic charges with a Mulliken charge analysis implemented in the generalized Bloch scheme. Fig.~\ref{charge} showcases the charge variation of atoms inside the structural unit located on the elongated side of the bent films, which reveals that electronic charges transfer from Mg to O atoms because these atoms are under the influence of tensional strains. The trend is more pronounced when the bending angles increase for both films.

\begin{figure}[tb]
\includegraphics[width=1\columnwidth]{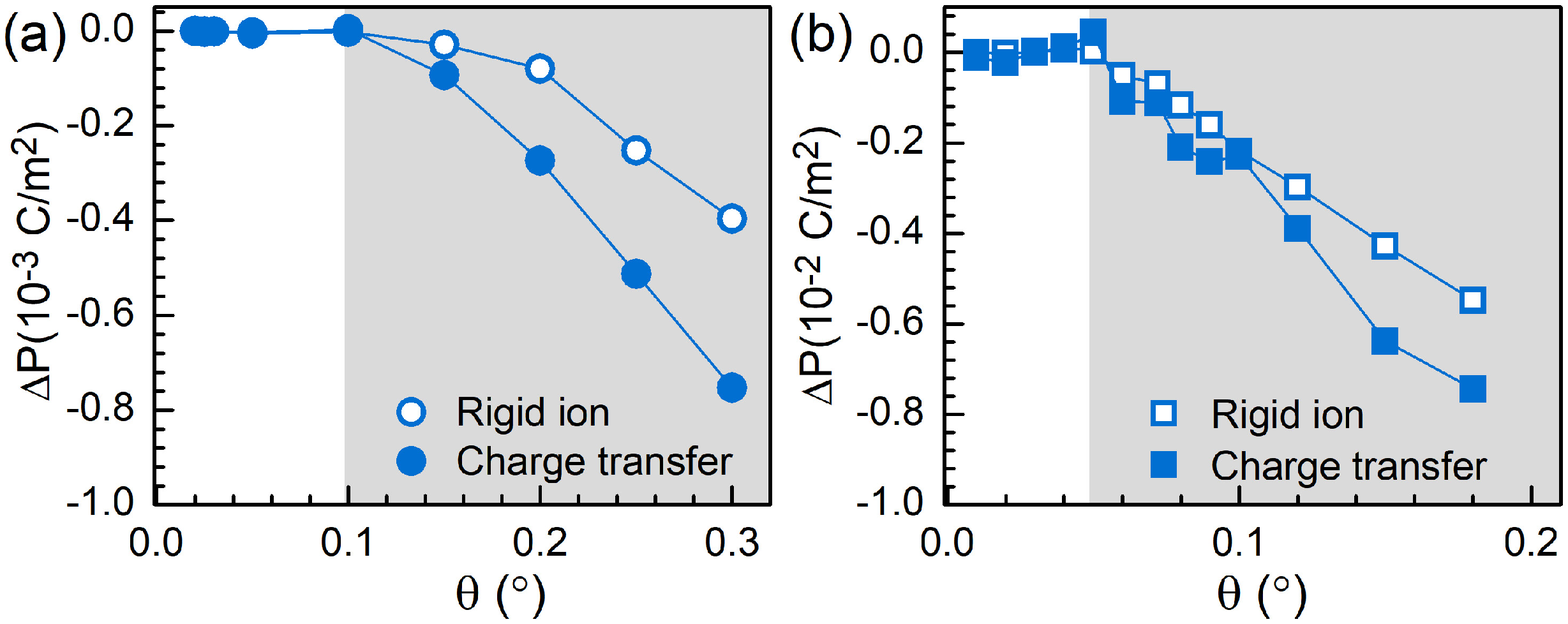}
\caption{Bending induced electric polarization of (a) (100) thinfilm and (b) (111) film. Both thinfilms are 10 nm thick.  }%
\label{polar}%
\end{figure}

The polarizations obtained considering both rigid-ion approximation and charge transfer are displayed in Fig.~\ref{polar}. Consistent with the above structural analysis, for both films, $\Delta {\bf P}\simeq 0$ in the linear elastic regime ($\theta<0.1^{\circ}$ for the (100) film and $\theta<0.04^{\circ}$ for the (111) film). On the other hand, in the nonlinear elastic regime ($\theta>0.1^{\circ}$ for the (100) film and $\theta>0.04^{\circ}$ for the (111) film), $\Delta {\bf P}$ becomes significant. Comparing the results considering the rigid-ion approximation and the results considering the charge transfer, the difference in $\Delta {\bf P}$ is non-negligible. This fact indicates that it is important to carry out QM simulations to obtain reliable polarizations. Using the data considering charge transfer, which displays roughly a linear dependence on the bending angle in the nonlinear elastic regime, we are able to determine the corresponding flexoelectric coefficients to be $\mu=0.09$~nC/m for the (100) thinfilm and $\mu=1.75$~nC/m for the (111) thinfilm.

To conclude, using the generalized Bloch theorem coupled with DFTB method, we investigate the flexoelectric effects of ions of MgO thinfilms under realistic bending deformations. Mechanically, for both the (100) and (111) films, their structural distortions with bending explicitly exhibit a linear elastic behavior at small bending angle and a nonlinear elastic behavior at large bending angle. Significant polarizations from ions are obtained in the nonlinear elastic regime where the strain gradient is high. This is because that only in this regime, atoms adopt discernibly large internal displacements. More, the internal displacements of atoms of the (111) film are much larger than those of the (100) film. Consequently, the obtained flexoelectric coefficient of the (111) film is also greater than that of the (100) film. These  reveal the anisotropic behavior of MgO film. Although the present study focuses on MgO thinfilm, we believe the flexoelectric effect can be found in many commonly-used semiconductors, and offers us a new way to engineer its electronic structure and construct a new-type semiconductor flexible electronic devices.

The authors thanks Jiawang Hong for insightful discussions. This work was supported by  the MOST of China (Grants No. 2016YFE0110000 and No. 2017YFA0303400) and  NSFC under Grants Nos. 11674022, 11874088  and U1930402. D.-B.Z. was supported by the Fundamental Research Funds for the Central Universities. Computations were performed at the Beijing Computational Science Research Center and Beijing Normal University.

%\bibliography{references}
%

\end{document}